\documentclass[12pt,preprint]{aastex}

\shorttitle{Classical Cepheid Pulsation Models}
\shortauthors{Petroni et al.}

\begin{document}

\title{Classical Cepheid Pulsation Models: IX. New Input Physics}

\author{Silvia Petroni}
\affil{Dipartimento di Fisica, Universit\`a degli Studi di Pisa, Via
Buonarroti 2, Pisa, 56127 Italy}
\email{petroni@df.unipi.it}

\author{Giuseppe Bono}
\affil{INAF-Osservatorio Astronomico di Roma, Via Frascati 33, 00040
Monte Porzio Catone, Italy} 
\email{bono@mporzio.astro.it}

\author{Marcella Marconi} 
\affil{INAF-Osservatorio Astronomico di Capodimonte, Via Moiariello 16, 
80131 Napoli, Italy}
\email{marcella@na.astro.it}

\author{Robert F. Stellingwerf} 
\affil{SC, 2229 Loma Linda, Los Alamos, NM 87544, USA}
\email{rfs@stellingwerf.com}


\begin{abstract}
We constructed several sequences of classical Cepheid envelope
models at solar chemical composition ($Y=0.28, Z=0.02$) to
investigate the dependence of the pulsation properties predicted by
linear and nonlinear hydrodynamical models on input physics. To study
the dependence on the equation of state (EOS) we performed several
numerical experiments by using the simplified analytical EOS
originally developed by Stellingwerf and the recent analytical EOS
developed by Irwin. Current findings suggest that the pulsation
amplitudes as well as the topology of the instability strip marginally
depend on the adopted EOS.

To compromise between accuracy and numerical complexity we computed
new EOS tables using the Irwin analytical EOS. We found that the
difference between analytical and tabular thermodynamic quantities and
their derivatives are smaller than 2\,\% when adopting suitable steps in
temperature and density.  To improve the numerical accuracy of
physical quantities we are now adopting bicubic splines to interpolate
both opacity and EOS tables. The new approach presents the substantial
advantage to avoid numerical derivatives both in linear and in
nonlinear models. The EOS first and second order derivatives are
estimated by means of the analytical EOS or by means of analytical
derivatives of the interpolating function. The opacity first order
derivatives are evaluated by means of analytical derivatives of the
interpolating function.

We also investigated the dependence of observables predicted by
theoretical models on the mass-luminosity (ML) relation and on the
spatial resolution across the Hydrogen and the Helium partial
ionization regions. We found that nonlinear models are marginally
affected by these physical and numerical assumptions. In particular,
the difference between new and old models in the location as well as
in the temperature width of the instability strip is on average
smaller than 200~K. However, the spatial resolution somehow affects
the pulsation properties. The new fine models predict a period at the
center of the Hertzsprung Progression ($P_{HP}=9.65$$-$9.84\, days) that
reasonably agree with empirical data based on light
curves ($P_{HP}=10.0\pm 0.5$\, days; \citealt{mbm92}) and on radial
velocity curves ($P_{HP}=9.95\pm 0.05$\, days; \citealt{mall00}), and
improve previous predictions by Bono, Castellani, and Marconi (2000,
hereinafter BCM00).
\end{abstract} 

\keywords{Cepheids -- Galaxy: stellar content -- hydrodynamics -- 
stars: evolution -- stars: oscillations }

\clearpage 
\section{Introduction}

The EOS and the opacity are fundamental physical ingredients for both 
evolutionary and pulsation models.  In particular, hydrodynamical models 
of variable stars, when compared with
static stellar structures, do require additional derivatives of
thermodynamic quantities (\citealt{c69}; \citealt{s74}, 1982, and
references therein).  As a consequence, the EOS, the opacities, and  
the numerical methods adopted to estimate physical quantities
such as pressure, temperature, internal energy, molecular weight, and
their derivatives, are crucial to properly compute the physical
structure of stellar envelopes \citep{dip91}.  Recent helioseismic
data uncorked several theoretical investigations aimed at improving
the accuracy of input physics currently adopted to construct both
solar and stellar models (see e.g. \citealt{call96,dall97,call02}).  The
EOS is typically provided in tabular form, where thermodynamic
quantities are provided for each given chemical composition as a
function of temperature and density. The most popular ones are the
so-called Mihalas-D\"{a}ppen-Hummer (MHD) EOS
\citep{mdh88,dall88,dam87} and the OPAL EOS developed at Livermore
(Rogers 1986; Iglesias and Rogers 1995; Rogers, Swenson, and Iglesias
1996) and recently improved in the treatment of input physics
\citep{r00,r01,rn02}.

To overcome the problems introduced by the interpolation across the
tables, new EOSs have also been developed in the form of in-line
analytical formulae that allow the estimate of thermodynamic
quantities as a function of temperature, density, and chemical
composition. The most recent ones are: EFF \citep{eff73}, CEFF
(Christensen-Dalsgaard and D\"{a}ppen, 1992),
Swenson-Irwin-Rogers-Eggleton-Faulkner-Flannery (SIREFF; see, e.g.,
\citealt{gs97}), and the recent in-line EOS developed by
Irwin\footnote{See also ftp://astroftp.phys.uvic.ca/pub/irwin/eos}
(\citealt{csi03}; \citealt{isvr03}).  The reader interested in a
detailed discussion concerning pros and cons of the different physical
assumptions adopted to derive these EOSs is referred to the reviews 
by D\"{a}ppen, Keady, and Rogers (1991), Rogers and Iglesias (1998), 
and by D\"{a}ppen and Guzik (2000).

Detailed numerical experiments concerning the dependence of pulsation
predictions on the EOS were performed by Kanbur (1991, 1992). He found
that both linear and nonlinear radiative Bump Cepheid models
constructed by adopting either the MHD EOS or the canonical Saha EOS
present negligible differences. However, more recent EOS calculations,
when compared with previous ones, do include a more detailed treatment
of heavy elements and of the most important Hydrogen molecules. As a
consequence, we decided to investigate the dependence of current
pulsation predictions for classical Cepheids on the input physics, and
in particular on the EOS by taking advantage of the analytical EOS
developed by Irwin. Moreover, recent theoretical investigations bring  
forward that the Mass-Luminosity (ML) relation might play a
fundamental role to account for actual properties of variable stars
\citep{bcm02,ball02}. Therefore, we also plan to test the
dependence on this key ingredient as well as on the spatial resolution
across the H and the He ionization regions.  The theoretical framework
adopted for constructing linear radiative and nonlinear, convective
models has already been described in a series of papers
(\citealt{bs94}; \citealt{bms99}, hereinafter BMS99; BCM00, and
references therein). The treatment of the EOS is described in
Stellingwerf (1975, 1982), while the treatment of the opacities is
discussed in \citet{bim96}.

In section~\ref{phys} we present the method adopted for handling the
opacity tables together with the comparison between the old EOS by
Stellingwerf and the new one by Irwin.  In section~\ref{lin} and
\ref{nonlin} we present detailed theoretical investigations aimed at
testing the dependence of theoretical observables predicted by
hydrodynamical models for Galactic Cepheids on the input physics. To
investigate the linear pulsation behavior we constructed a set of
nonadiabatic radiative models. The linear observables, namely period
and blue boundary of the instability strip, are widely discussed in
section \ref{lin1} and \ref{lin2}. We also constructed nonlinear and
time-dependent convective Cepheid models to assess the modal stability 
and, in turn, to evaluate both the boundaries of the strip as well as 
the amplitude and the morphology of light and velocity curves (section 
\ref{nonlin1} and \ref{nonlin2}). Theoretical predictions are also 
compared with observational data available in the literature. Finally,
section \ref{hp} is focused on the Hertzsprung-Progression (HP). A
summary of the results is given in section \ref{summa} together with 
a brief discussion concerning future plans.

\section{Input physics} \label{phys}

\subsection{The opacity} \label{opacity}

The pulsation models are constructed by adopting radiative opacities
(OPAL; \citealt{ri92}, \citealt{ir96}) for temperatures higher than
$\sim 6000$~K and molecular opacities (\citealt{af94}, hereinafter AF)
at lower temperatures. The method adopted to handle the opacity
tables, and to evaluate the opacity derivatives with respect to
temperature and density, is a revised version of the method described
by Bono, Incerpi, and Marconi (1996). It relies on bicubic
interpolating functions with analytical derivatives whose coefficients
are computed from the functions and the derivatives at grid points
\citep{s93}.  In the previous approach, for each given chemical
composition, two distinct programs were developed to compute a finer
opacity table by interpolating the original OPAL and AF
tables. Together with the opacity, the opacity derivatives with
respect to temperature and density were also calculated and stored.
The previous tables were matched together for a temperature equal to
10,000~K.  Finally, the opacity and its derivatives were computed by
means of a bilinear interpolation on the finer big table.

The revised method only relies on bicubic spline interpolations.  The
original two sets of opacity tables (OPAL and AF) are first matched
together, then the bicubic spline and its derivatives are computed
directly inside the pulsation codes, for the entire range of
temperature and density covered by the original table. The temperature
value we adopted to match the two sets of opacities is $T=6300$~K
\citep{vsria00}, since in this temperature range the opacity
variations are smooth.  Moreover, at lower temperatures, the density
range has been extended from $\log R=1.0$ up to $\log R=7.0$, where 
$R=\rho/{T_{6}}^{3}$, ${T_{6}}=10^{-6}$ $T$, and $\rho$ is the density.
The new method allowed us to decrease by approximately a factor of ten
the CPU time required to construct nonlinear models.

\begin{figure}
\epsscale{0.80}
\plotone{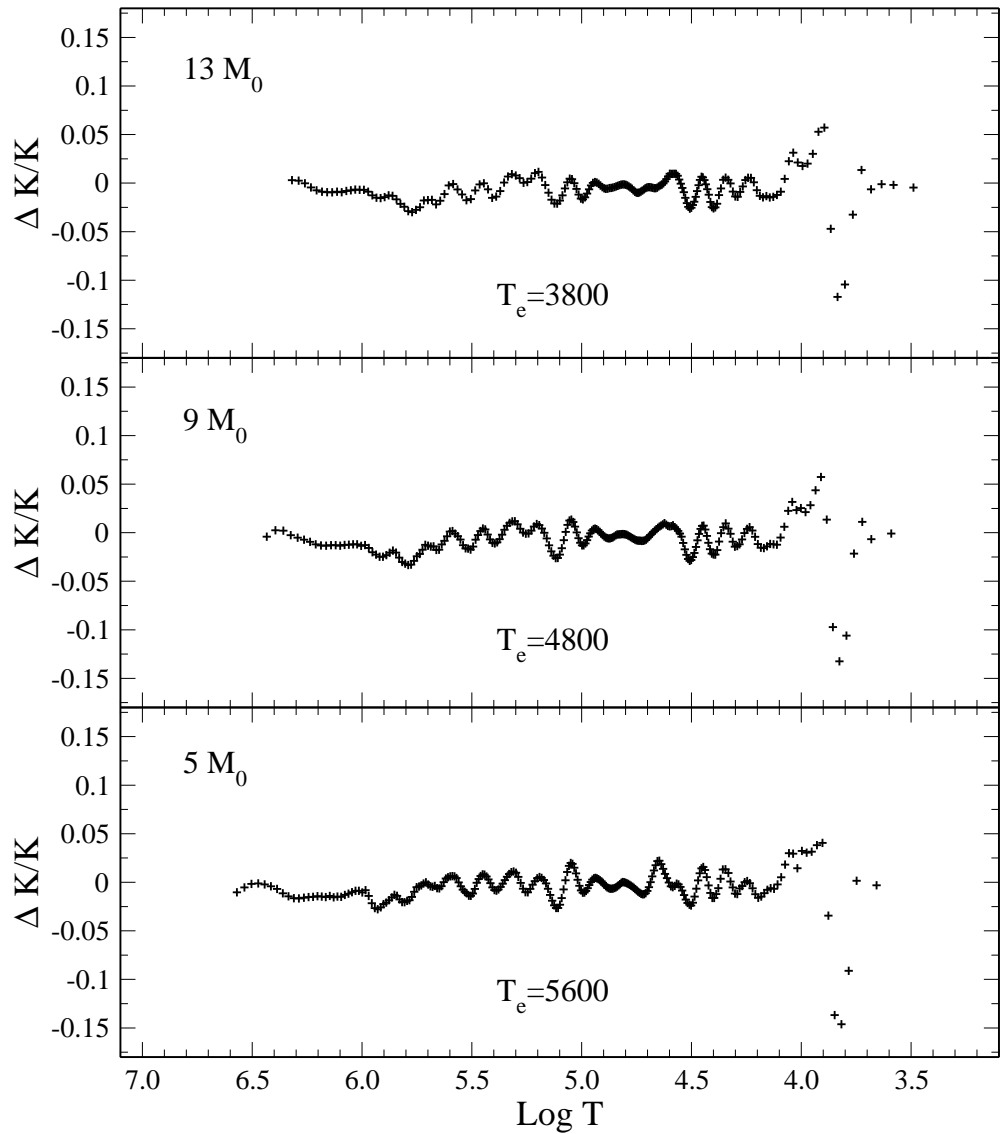}
\caption{Relative difference in opacity between the old and the new
method adopted to interpolate opacity tables as a function of
logarithmic temperature.  From top to bottom data refer to different
linear models. See text for more details.\label{fk}}
\end{figure}

To investigate whether the method adopted to interpolate the opacity
tables affects the structure of pulsating models, we performed several
numerical experiments across the instability strip. Figure \ref{fk}
shows the relative difference in opacity between the old and the new
interpolation method for three Cepheid models at solar chemical
composition, namely $Y=0.28$, and $Z=0.02$. The adopted stellar mass
ranges from 5~$M_{\sun}$ to 13~$M_{\sun}$, while the effective
temperature ranges from $T_e=5600$~K to $T_e=3800$~K.  The difference
is on average smaller than 1\,\% and approaches 2\,\% at $\log
T\approx 3.8-4.0$~K, i.e. the region where the radiative (OPAL) and
the molecular (AF) opacity tables were matched together (6300 vs
10,000~K). This indicates that the two methods approach the accuracy
of opacity tables that is typically of the order of 2\,\% \citep{s93}.
Note that to avoid spurious wiggles in the comparison between the two
interpolation methods due to small changes in the zoning of the models
(Guzik and Swenson 1997), we computed at first the models with the old
method and then we used the same values of temperature and density of
the individual models to interpolate the opacity tables with the new
method.  The marginal difference between the two interpolation methods
is further supported by the fact that two independent sets of linear
models constructed by adopting the old and the new interpolation
method present a negligible difference both in the pulsation period
and in the growth rate (see data listed in the top of Table~\ref{t1}).

\subsection{The equation of state} \label{eos}

Pulsation models constructed by our group rely on the equation 
of state developed by Stellingwerf (1975, 1982). It treats equilibrium
mixture of $H$, $H^{+}$, $He$, ${He}^{+}$, ${He}^{++}$, $M$, and
$M^{+}$, where $M$ designates a fictional {\em metal} with fixed
ionization potential, number abundance, atomic weight, and degeneracy
ratio. These parameters were chosen by fitting the electron pressure
at cool temperatures given by an exact solution of King IVa ($Y=0.28$, 
$Z=0.02$) chemical composition \citep{ckt73}. This fictional element 
was included to account for the ionization of $Mg$, $Fe$, and $Si$ in 
cooler Cepheid envelopes \citep{s82}.

The accuracy in opacity and EOS derivatives is a key feature to
construct accurate pulsation models, since the driving mechanisms are
connected with envelope regions where the temperature and the density
gradients present sudden changes. Therefore, we decided to improve our
hydrodynamical codes by adopting the analytical EOS developed by
Irwin, since it is based on approximately the same physics adopted in
recent EOS such as MHD and OPAL, and it is quite flexible. We adopted
the option suggested by Irwin ({\em EOS1}) that was constrained by
fitting the OPAL and the SCVH \citep{scv95} EOSs.  The high numerical
complexity typical of in-line routines was partially overcome, by
selecting some input parameters that allow a sound compromise between
speed and accuracy.  The flexibility is the main advantage in using an
analytical EOS, since it can cope with variations in chemical
composition and in elemental mixture. Moreover, an analytical EOS
provides smooth high-order derivatives of physical quantities and it
can also be easily modified to compute additional thermodynamic
quantities.  On the other hand, the tabular EOS requires a limited
amount of CPU time to interpolate the physical quantities when
compared with the in-line EOS. Although, the tabular EOS might be
limited in the $\log T$ vs $\log \rho$ coverage as well as in chemical
compositions, and does not include high-order derivatives of
thermodynamic quantities (Guzik and Swenson 1997; D\"{a}ppen and Guzik
2000). The latter problem is generally overcome by computing numerical
derivatives.

Keeping in mind these problems, we decided to investigate the key
features of EOSs available in the literature. The Irwin EOS includes
neutral and positive ions of the 20 most abundant atoms, namely: $H$,
$He$, $C$, $N$, $O$, $Ne$, $Na$, $Mg$, $Al$, $Si$, $P$, $S$, $Cl$,
$A$, $Ca$, $Ti$, $Cr$, $Mn$, $Fe$, $Ni$. On the other hand, the
tabular EOS by MHD and OPAL only include the elements $H$, $He$, $C$,
$N$, $O$, and $Ne$, with the last one that is considered as
representative of heavier elements \citep{rsi96,rn02}.  The Irwin EOS
allows the use of different solar element mixtures, and therefore the
possibility to use the same mixture adopted in the calculation of
opacity tables.  Note that an EOS that accounts for molecular species
might play an important role in constructing pulsation models for
long-period-variables such as Semi-Regular, Miras and long-period
Cepheids.  The OPAL project recently provided new EOS tables that
include $H_2$, $H_2^+$ as well as $H^-$, $He_2^+$, and $HeH^+$
\citep{rn02}, while the Irwin EOS only includes the hydrogen
molecules.  However, previous authors\footnote{See also
http://www-phys.llnl.gov/Research/OPAL/index.html} found that 
in physical regimes where the EOS shows rapid changes, and the 
density is lower than $\log \rho < -8$, typical of Cepheid envelopes 
at temperatures cooler than 60,000~K, first order properties, such as
pressure and internal energy, might be off by as much as a few
percents, while second order properties such as specific heat and
adiabatic exponents might be off by as much as 10\,\%.

\begin{figure}
\epsscale{0.80}
\plotone{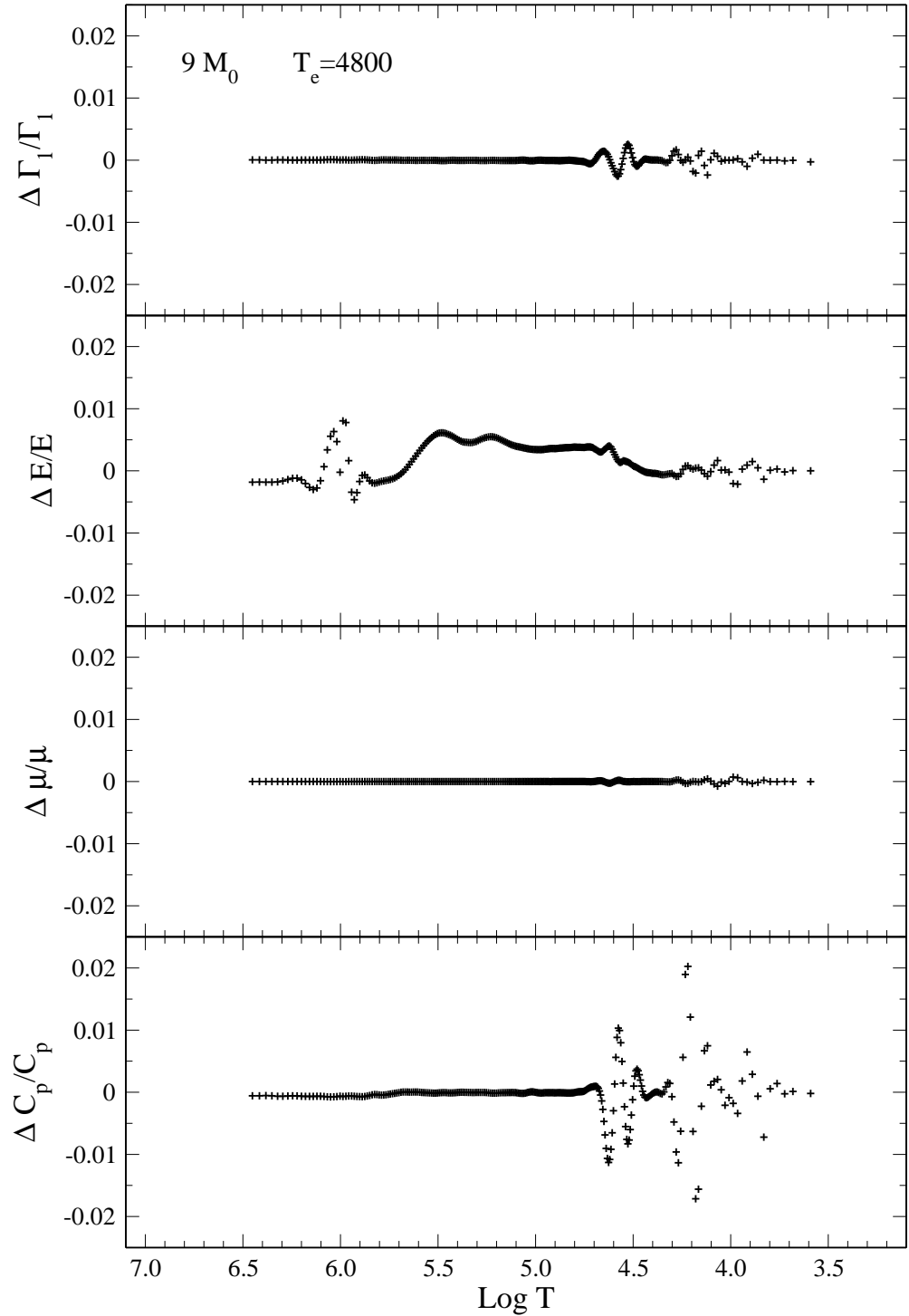}
\caption{From top to bottom relative difference in adiabatic exponent,
internal energy, molecular weight, and specific heat at constant
pressure between the analytical and the tabular form of the Irwin
EOS.\label{feosI}}
\end{figure}

In this context, it is worth noting that the main difference in
dealing with linear and nonlinear pulsation models is that the latter
ones require the second derivatives of total pressure (see eq. A21-A23
in \citealt{s82}). These quantities are not generally included in
tabular EOSs, and to compute them it is necessary to perform numerical
derivatives across the tables. This means that the second derivatives
are likely less accurate than the other thermodynamic quantities.  To
compromise between accuracy and numerical complexity, we decided to
use the Irwin EOS to compute several tabular EOS at fixed chemical
composition but with different grid spacings. The resolution in
temperature and density of the tabular EOS need to be cautiously
treated, since the interpolation across the tables or the estimate of
numerical derivatives might be affected by systematic uncertainties if
the grid resolution is too coarse \citep{dip91}. We performed several
numerical experiments to overcome this problem, and we found that EOS
tables with a step ranging from $\sim 0.05$ dex to $\sim 0.1$ dex in
$\log T$ and 0.5 dex in $\log R$ works quite well. Figure \ref{feosI}
shows the relative difference in four physical quantities between the
analytical and the tabular EOS by Irwin for a 9~$M_{\sun}$ model with
$T_e = 4800$ K. From top to bottom the figure shows the relative
difference in the adiabatic exponent $\Gamma_{1}$ \citep{cg68},
internal energy $E$, mean molecular weight per particle $\mu$, and
specific heat at constant pressure $C_{p}$.  Data plotted in this
figure suggest that the difference is typically smaller than 1\,\%,
while for $C_p$ it attains the 2\,\% across the Hydrogen Ionization
Region (HIR). Note that, to perform the difference, we constructed at
first the linear model using the tabular EOS and then we used the same
values of temperature and density to estimate with the analytical EOS
the same physical quantities. Current finding is further supported by
the fact that the difference in pulsation periods and growth rates
between linear Cepheid models constructed by adopting the analytical
EOS and the tabular EOS is at most of the order of a few thousandths
(0.003, see data listed in the middle of Table~\ref{t1}).

We performed the same test using a wide range of stellar input
parameters and the outcome was the same. This means that the current
approach allow us to derive thermodynamic quantities and their
derivatives with an accuracy that is on average better than 2\,\%. The
method adopted for interpolating tabular EOS and to derive high-order
derivatives is the same adopted for the opacity tables (see Section
\ref{opacity}).  Note that current method uses bicubic splines that
provide the opportunity to derive analytical first and second order
derivatives instead of estimating them numerically.  As a matter of
fact, we computed with the analytical Irwin EOS the first derivatives
of pressure with respect to temperature and density, then the
interpolation with the bicubic splines allowed us to derive
analytically the three second-derivatives of pressure. At the same
time, we computed with the analytical EOS the specific heat at
constant pressure and the interpolation provided the analytical
first-derivatives with respect to temperature and density.

\section{Linear results} \label{lin}

\subsection{Dependence on the EOS} \label{lin1}  

To investigate the dependence of pulsation properties on the EOS, we
constructed a survey of linear, nonadiabatic models at solar chemical
composition, namely $Y=0.28$, $Z=0.02$. The luminosity of Cepheid
envelope models was fixed according to the ML relation recently 
derived by Bono et al. (2000). This relation relies on
several sets of evolutionary models and agrees quite well with similar
prescriptions available in the literature. 
The difference between the current ML relation and the ML relation
provided by \citet{abha99} and by \citet{ccs92} is, at fixed stellar
mass, on average smaller than 0.1 dex.
To constrain the effect of the EOS, we selected three Cepheid models
that cover a substantial portion of the instability strip. In
particular, the model with 5~$M_{\sun}$ is located close to the
fundamental blue edge ($T_e=5600$~K), the 9~$M_{\sun}$ model is
approximately located in the middle of the strip ($T_e=4800$~K), while
the 13~$M_{\sun}$ model is close to the red edge ($T_e=3800$~K).

Current numerical experiments show that the relative difference
in $\Gamma_{1}$ and in $C_{p}$ is generally smaller than the
uncertainties due to the interpolation scheme (see
Fig.~\ref{feosI}).  However toward lower temperatures, the
difference between the two EOSs increases and becomes of the order of
15\,\% ($\Gamma_{1}$) and 30\,\% ($C_{p}$) close to the surface of these
envelope structures. The relative difference in the mean molecular
weight, $\mu$, presents a systematic trend but it is vanishing
(smaller than 1\,\%) throughout the entire envelope.  On the other hand,
the difference in the internal energy attains vanishing values
(smaller than 1\,\%) over a substantial portion of the envelope, it
undergoes a sudden increase in the outermost regions ($T\le 10,000$~K),
and approaches a difference of the order of 50-60\,\% close to the
surface. Note that to avoid spurious wiggles in the comparison between
the two EOSs, we constructed at first the models with the analytical
Stellingwerf EOS and then we used the same values of temperature and
density to interpolate the tabular Irwin EOS.

Even though some physical quantities present a sizable difference
between the Stellingwerf and the Irwin EOS, the impact of this
difference on linear observables such as pulsation periods and growth
rates is negligible. The difference in the linear periods
(fundamental, first, and second overtone) is at most a few hundredths
(see data listed in the bottom of Table~\ref{t1}). 
The same outcome applies for the growth rates. The negligible difference
between the two sets of models is due to the marginal effect that the
outermost layers have on the pulsation properties.

\subsection{Blue boundaries} \label{lin2}  

To investigate the dependence of the blue (hot) edges of the Cepheid
instability strip on the EOS, we constructed several sequences of
models with stellar masses ranging from 5 to 13~$M_{\sun}$. As in
previous investigations (BMS99), the adopted effective temperature
step is 100~K. Current ML relation predicts, for each given stellar
mass, a luminosity level that is on average lower than adopted by
BMS99 and by BCM00. The difference in luminosity is vanishing in the
low-mass range but increases when moving toward higher stellar masses.
The main intrinsic properties of hydrostatic envelopes and the basic
assumption on the numerical approximations adopted for constructing
the models have already been discussed in a series of previous papers
(see, e.g., \citealt{bs94}; BMS99; BCM00). The new models, when
compared with the old ones, present a finer spatial resolution in the
region located between the first Helium Ionization Region (HeIR,
$T\approx 2.1\times 10^4$~K) and the surface. The number of zones in
this region is typically 35, i.e. $\Delta T\approx 450 - 520$~K, while
in the old ones was $\approx25$, i.e. $\Delta T\approx 600 -
680$~K. The inner boundary condition was fixed in such a way that the
base of the envelope is located at a distance
$r_{0}=r/R_{ph}=0.03-0.04$ from the star center, where $R_{ph}$ is the
equilibrium photospheric radius, while in the old models it was fixed
at $r_{0}=r/R_{ph}\approx0.1$. The envelope mass ranges from 65\,\% to
40\,\% of the total mass when moving from hotter (5~$M_{\sun}$) to
cooler (13~$M_{\sun}$) Cepheid models.

To investigate the dependence of the blue edges on the equation of
state, we constructed two independent sets of linear models according
to the Stellingwerf and the Irwin EOSs.  Current numerical
experiments show that the location of fundamental (F) and
first-overtone (FO) linear blue boundaries mildly depend on the
adopted EOS.  The F edges based on the Irwin EOS either are identical, 
in the low-mass regime, to the edges based on the Stellingwerf EOS
or they present a difference at most of 200~K for $M/M_{\sun}=9$. The
difference for FO edges is even smaller and at most of $\sim 100$~K. 
However, it turns out that the spatial resolution and the ML
relation affect the linear blue boundaries more than the EOS. In fact,
the new F blue edge is on average 200~K hotter than predicted by
BMS99, and the difference becomes $\sim 400$~K for stellar masses
ranging from 7 to 9~$M_{\sun}$. Note that the difference in the
predicted luminosity for $M/M_{\sun}=5$ between Bono et al. (2000) and
BMS99 ML relations is vanishing. Therefore, the difference of 300~K
between new and old F and FO blue edges is due to the increase in the
spatial resolution of the new models.

\section{Nonlinear results} \label{nonlin}

\subsection{The instability strip} \label{nonlin1}

Nonlinear and time-dependent convective hydrodynamical models supply
the unique opportunity to estimate both the blue and the red edge of
the instability strip as well as to provide robust predictions
concerning the pulsation amplitudes, and the shape of light and
velocity curves along the pulsation cycle. This means that observables
predicted by nonlinear models can be soundly compared with actual
properties of variable stars. To supply a new homogeneous scenario, we
constructed six new sequences of nonlinear models with stellar masses
ranging from 5 to 13~$M_{\sun}$.  The nonlinear analysis was performed
by perturbing the linear F and FO radial eigenfunctions with a
velocity amplitude of 5~Km~s$^{-1}$.  Table~\ref{t2} summarizes the
input parameters as well as the nonlinear pulsation properties of the
entire set of models. The observables listed in this table have been
estimated at full amplitude, i.e. after that the perturbed envelopes
approached the nonlinear limit cycle stability (see BMS99 for more
details).

\begin{figure}
\epsscale{0.80}
\plotone{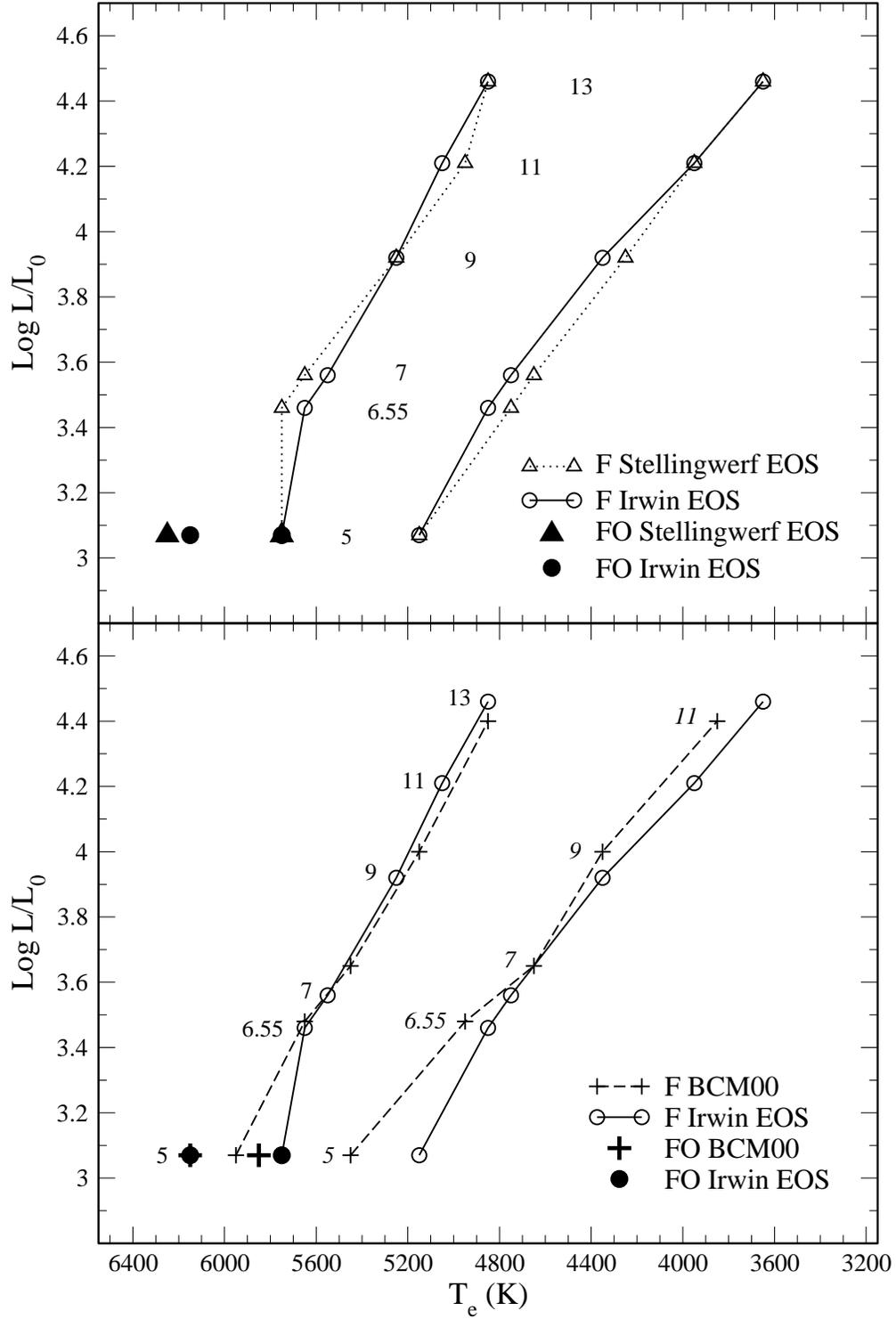}
\caption{Comparison in the HR diagram between nonlinear Fundamental
(F) and First Overtone (FO) blue and red edges according to models
constructed using different EOS (top panel) or different EOS, ML
relation, and spatial resolution (bottom panel). Stellar masses are in
solar units. \label{fnolin}}
\end{figure}

To investigate the dependence of nonlinear pulsation models on
the EOS we also computed six sequences of pulsation models by adopting
the same assumptions and input parameters of the models listed in
Table~\ref{t2}, but by adopting the Stellingwerf EOS. The observables
predicted by these models are given in Table~\ref{t3}\footnote{This 
Table is only available in the on-line edition of the manuscript.}. 
The top panel of Figure \ref{fnolin} shows the comparison between 
nonlinear F and FO
edges predicted by models that use the Stellingwerf (triangles) or the
Irwin (circles) EOSs. Once again data plotted in this panel disclose
that the difference is smaller than 100~K. The same outcome applies
for the temperature width, and indeed it decreases by 200~K for
stellar masses across the so-called Hertzsprung
Progression\footnote{In the period range from 6 to 16 days classical
Cepheids present a well-defined Bump along both the luminosity and the
radial velocity curve. For periods shorter than $\approx 9$ days the
Bump is located along the decreasing branch, while toward longer
periods it takes place along the rising branch.} (HP), i.e. $M\approx
6.55, 7 \, M_{\sun}$, while for the other mass values the difference
is $\le 100$~K.  Moreover, data listed in Tables~\ref{t2} and \ref{t3}
show that the relative difference in periods between models
constructed by adopting the Stellingwerf EOS and the models based on
the Irwin tabular EOS is smaller than 1\,\% across the entire
instability strip.  These findings, together with the results obtained
for the linear blue edges, indicate that the EOS has a marginal effect
on the topology of the Cepheid instability strip.

To single out the effect of the ML relation on pulsation properties we
also constructed a new sequence of nonlinear models for $M=11 \
M_{\sun}$ and $\log L/L_{\sun}=4.40$ (see Table~\ref{t4}). These
models when compared with the sequence of models for $M=11 \, M_{\sun}$
listed in Table~\ref{t2}, only differ in the luminosity level (4.40
[BCM00] vs 4.21 [Bono et al. 2000]). The comparison between the
observables given in Table~\ref{t2} and \ref{t4} shows that the
dependence of the instability strip on the ML relation is mild. The
shift in temperature of the new edges is typically smaller than 200 K,
while the pulsation amplitudes attain quite similar values.

To investigate the dependence of the pulsation behavior on the spatial
resolution across the HIR, we constructed a new sequence of nonlinear
models for $M=11 \, M_{\sun}$ by adopting the same input parameters of
models listed in Table~\ref{t2}, but a coarse zoning (see
Table~\ref{t5}). Data listed in Tables~\ref{t2} and \ref{t5} disclose
that the spatial resolution affects the modal stability, and indeed
the F edges shift by $\approx 200$~K toward hotter effective
temperatures. This finding is also supported by the set of models for
$M=5 \, M_{\sun}$ listed in Table~\ref{t3}. These models and the set
of models for $M=5 \ M_{\sun}$ constructed by BCM00 present the same
luminosity, but the comparison displays a shift of $200-300$~K in the
F edges. Moreover, finer models present an increase in the temperature 
width of the instability region, namely 1000 vs 600~K for the $M=5\,M_{\sun}$ 
models and 900 vs 700~K for the $M=11\, M_{\sun}$ models. At the same 
time, the region where FO are pulsationally unstable increases 
(500 vs 300~K), while the difference in the temperature width for 
F pulsators is marginal (600 vs 500~K). Note that the increase in the
spatial resolution also affects the pulsation amplitudes, and indeed
the fine zoning models present, at fixed distance from the blue edge,
bolometric amplitudes that are 10\,\% larger than the coarse zoning
models. This difference is even greater ($\sim 25\,\%$) for the $
M=5\,M_{\sun}$ FO models.

The bottom panel of Figure \ref{fnolin} shows the comparison between
the new F and FO edges (circles) and the edges predicted by BCM00
(pluses). A glance at the data plotted in this panel shows that the
topology of the instability strip predicted by old and new models is
quite similar.

Figures 4, 5, 6, 7, 8, and 9 display both light and velocity curves as
a function of phase over two consecutive pulsation cycles for the
entire set of models\footnote{The figures 4, 7, 8, and 9, as well as
figures 12 and 13 are only available in the on-line edition of the
manuscript.}. The comparison between these figures and predictions
obtained for the set of models listed in Table~\ref{t3} further
strengthens the evidence that the EOS also marginal affects the
morphology of light and velocity curves.

\subsection{Pulsational amplitudes} \label{nonlin2}

Nonlinear models also predict the luminosity amplitude, that is a key
parameter to study the pulsational properties of classical Cepheids.
In particular the Bailey Diagram, i.e. luminosity amplitude versus
period, is only marginally affected by reddening and distance
uncertainties. Therefore, the comparison between theory and
observations in this plane supplies independent constraints on the
intrinsic Cepheid parameters to be compared with evolutionary
prescriptions.  The top panel of Figure \ref{fabolp} shows predicted
bolometric amplitudes of Cepheid models constructed by adopting the
new (solid line) and the old (dotted line) EOS as a function of
period. The agreement between the two sets of models is very good over
the entire mass range.  The first overtone models for $M=5 \,
M_{\sun}$ and the fundamental models across the HP present a mild
difference on the EOS close to the edges of the instability region.
Data plotted in the middle panel show the dependence of bolometric
amplitudes on ML relation and spatial resolution when the Stellingwerf
EOS is adopted in the computations. As a whole, the bottom panel
displays the comparison between new predictions and old results by
BCM00. The predicted luminosity amplitudes show a {\em hook} shape 
when moving from the blue to the red edge of the instability region, 
while the models located across the HP present the typical 
{\em double-peaked} distribution of Bump Cepheids (Bono, Marconi, 
and Stellingwerf 2000, hereinafter BMS00). This trend supports the 
empirical evidence for Galactic
Cepheids originally brought forward by \citet{st71} and by Cogan
(1980).  They found that in the period range from $\log P \approx
0.40$ to 0.86 and for $\log P > 1.1$-1.3 the largest luminosity
amplitudes are attained close to the blue edge, while across the HP
($0.85 \le \log P \le 1.1$-1.3) the maximum is attained close to
the red edge (BCM00, BMS00).

\subsection{Hertzsprung Progression} \label{hp}

\begin{figure}
\epsscale{0.80}
\plotone{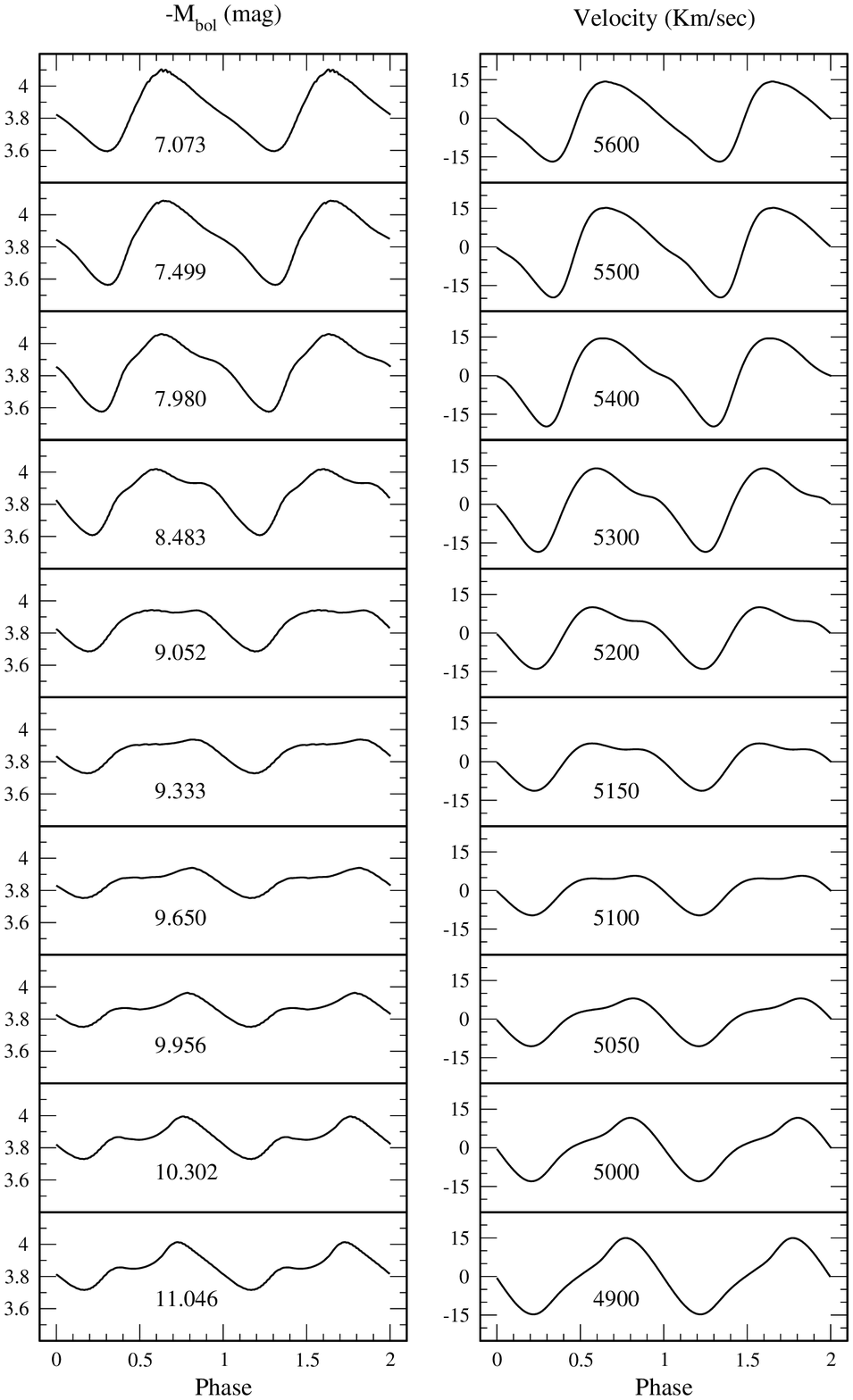}
\caption{Same as Fig. 4, but for the sequence of models with
$M=6.55 \, M_{\sun}$.
\label{fc655}}
\end{figure}

\begin{figure}
\epsscale{0.80}
\plotone{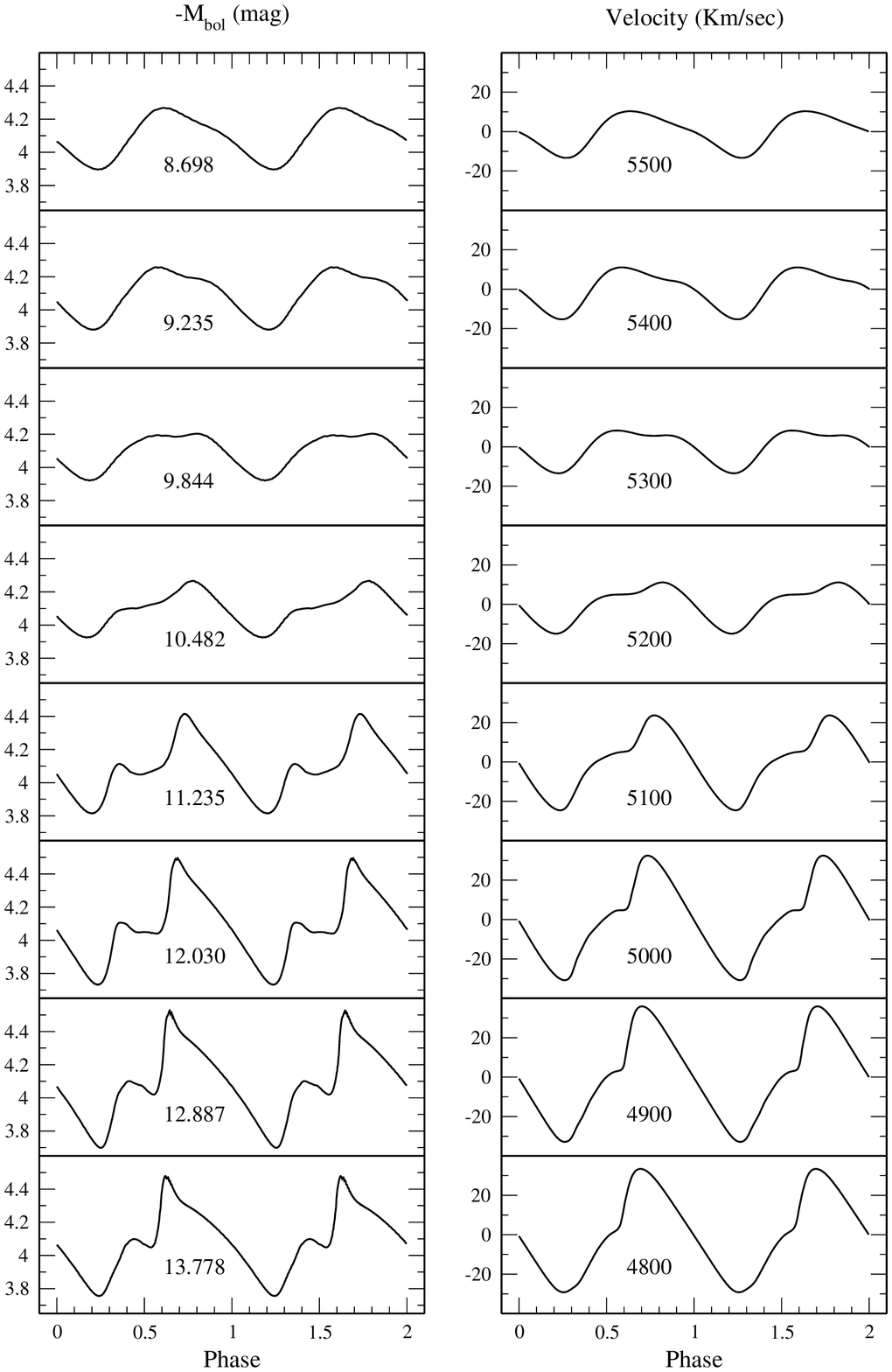}
\caption{Same as Fig. 4, but for the sequence of models with
$M=7 \, M_{\sun}$.
\label{fc7}}
\end{figure}

\begin{figure}
\epsscale{0.80}
\plotone{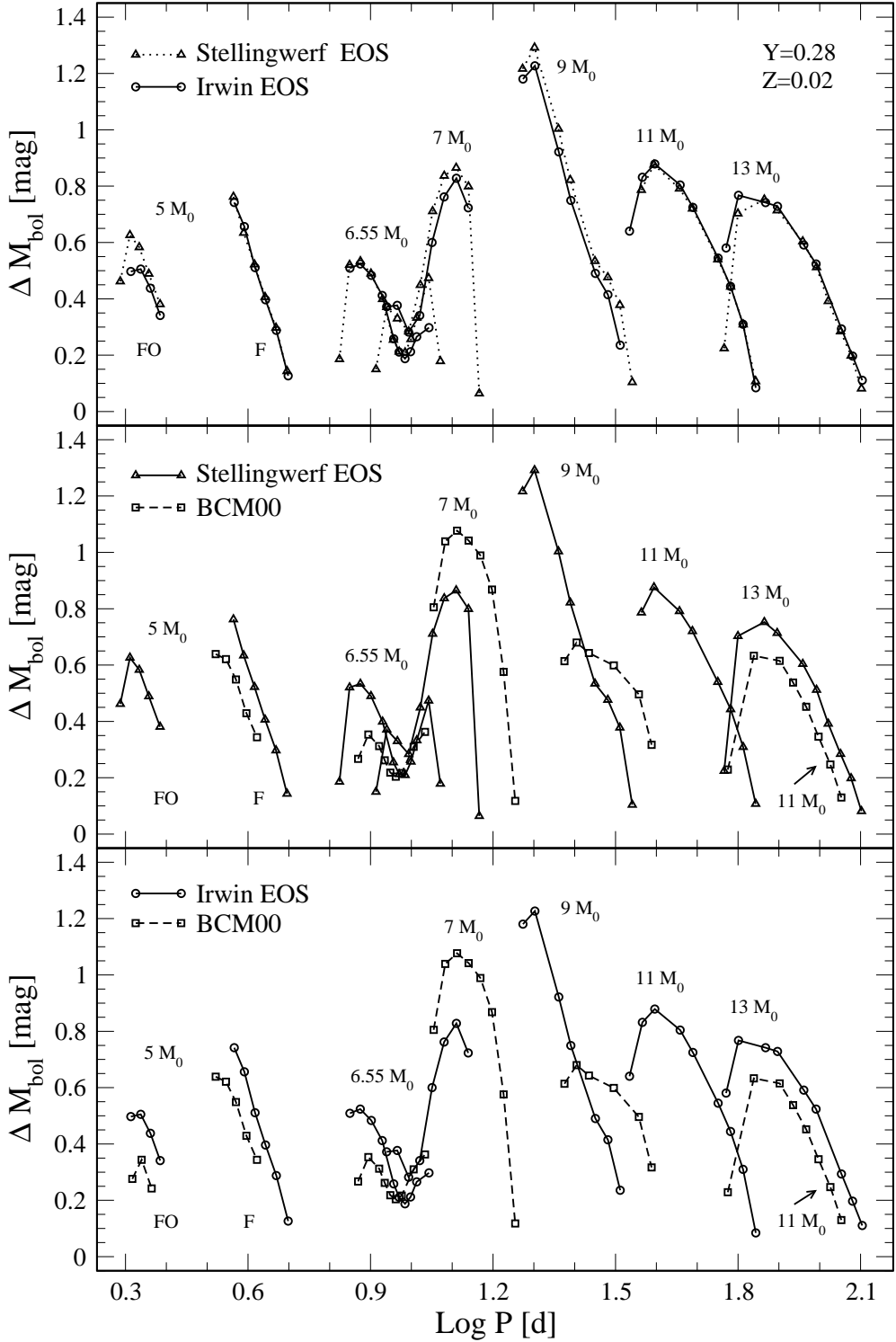}
\caption{Comparison between bolometric amplitudes based on pulsation
models constructed by adopting different EOS (top panel), different ML
relation, and spatial resolution (middle panel), different EOS, ML
relation, and spatial resolution (bottom panel).\label{fabolp}}
\end{figure}

\begin{figure}
\epsscale{0.80}
\plotone{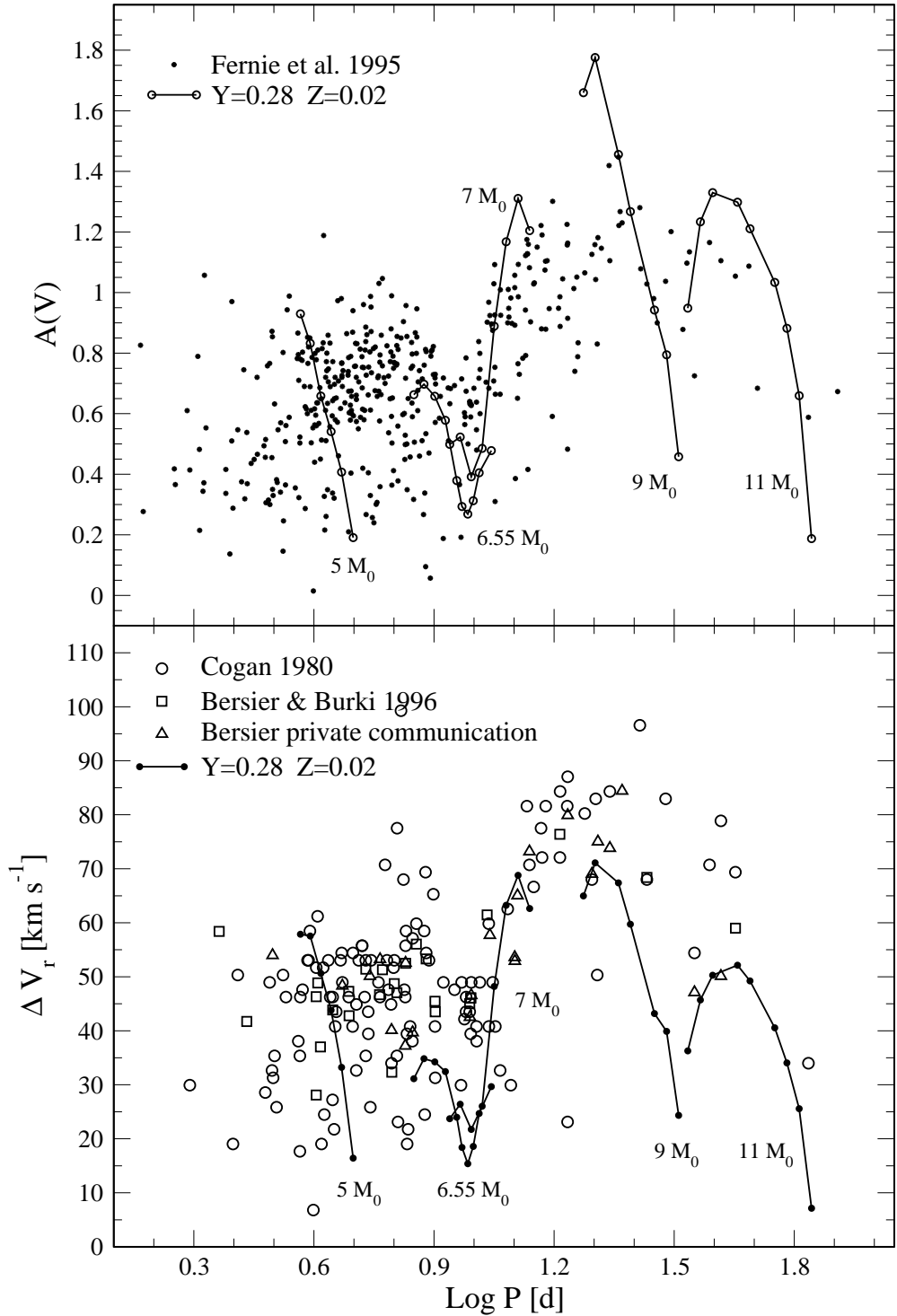}
\caption{Comparison between predicted and empirical amplitudes for
fundamental Galactic Cepheids. The top panel shows the comparison
between V-band amplitudes collected by Fernie et al. (1995) and
current models.  The bottom panel displays the comparison between
different samples of empirical radial velocity amplitudes and
theoretical predictions. Note that predicted amplitudes were
multiplied by 1.36 \citep{bb96}.\label{favp}}
\end{figure}

The top panel of Figure \ref{favp} shows the comparison between
predicted V-band amplitudes and empirical data for Galactic Cepheids
collected by Fernie et al. (1995), while the bottom panel displays the
comparison between predicted and empirical radial velocities. Theory
is in reasonable agreement with observations. The models across the HP
deserve a more detailed discussion.  According to BMS00 the secondary
minimum in the luminosity and in the velocity amplitude of classical
Cepheids located at $\log P\approx 1$ can be adopted to fix the period
of the HP center ($P_{HP}$). They found that for LMC Cepheids
$P_{HP}=11.24 \pm 0.46$\,days, while current models for Galactic
Cepheids suggest that it is located between $P_{HP}\approx 9.65$\,days
($M=6.55 \, M_{\sun}$, $T_e=5100$~K) and $P_{HP}\approx 9.84$\,days
($M=7 \, M_{\sun}$, $T_e=5300$~K). Current estimates are in good
agreement with empirical data, and indeed \citet{mbm92} found that the
minimum in the Fourier parameters of Galactic Cepheid light curves is
roughly equal to $P_{HP}=10.0\pm 0.5$\,days, and in reasonable
agreement with Moskalik et al. (2000) who found $P_{HP}=9.95\pm
0.05$\,days using radial velocity curves of 131 Cepheids. This finding
confirms the results obtained by BMS00, i.e. an increase in the metal
content causes a shift of the HP center toward shorter periods. In
fact, for SMC ($Z\approx0.004$) and LMC ($Z\approx0.008$) Bump
Cepheids, it is located at $11.0 \pm 0.5$ and $10.5 \pm 0.5$\,days
respectively \citep{b98}. Moreover, predicted luminosity and velocity
curves plotted in Figures \ref{fc655} and \ref{fc7} also disclose that
the bump along the light curves crosses the luminosity maximum at
shorter periods when compared with the velocity curves. A similar
feature was already found by BMS00 for LMC Cepheids. Light and
velocity curves displayed in Figure 12 refer to models
constructed by adopting the same input parameters of models in Figure
\ref{fc655} ($M = 6.55 \, M_{\sun}$, $\log L/L_{\sun} = 3.46$), but
the Stellingwerf EOS. Data plotted in this figure, together with data
listed in Table~\ref{t3}, show that the EOS also has a marginal effect
on Bump Cepheids, and indeed the changes along the pulsation cycle as
well as in the HP between the two sets of models are quite small.

To constrain the effect of the ML relation and of the spatial
resolution on the HP we computed a new sequence of models for $M =
6.55 \, M_{\sun}$ by adopting the same input physics and parameters of
the models by BCM00 (see Table~\ref{t6} and Figure 13).

Data plotted in Figure 12 and 13 are both based
on the Stellingwerf EOS, but differ for the luminosity level ($\log
L/L_{\sun}=3.46$ [Bono et al. 2000] vs 3.48 [BCM00]) and for the
spatial resolution, since models in Figure 13 are
characterized by a coarse zoning across the ionization regions.
Interestingly enough, the comparison between these two figures
discloses that models constructed by adopting the same input physics,
a quite similar luminosity level, but a different zoning causes a
shift of the HP center toward longer periods, and in turn a better
agreement with empirical data.  A glance at the data displayed in
Figure 13 shows that for $M=6.55 \, M_{\sun}$ the HP center
is, indeed, located at $P_{HP}\approx 9.16$\,days ($T_e\approx5250$).
Note that for $M=7 \, M_{\sun}$ the models constructed by BCM00 do not
show the HP (see their Fig. 11j in the on-line edition).


Qualitative arguments concerning the shape of the fundamental light
curves indicate that also the new models for $M=5 \, M_{\sun}$ show a
well-defined bump across the instability region.  To assess on a
quantitative basis whether theoretical models account for the HP that
has been detected among short-period Galactic Cepheids, it is
necessary to implement current models with new ones that cover the
low-mass range (Bono et al. 2002).

\section{Summary and final remarks} \label{summa}

We performed several numerical experiments aimed at testing the
dependence of pulsation observables predicted by both linear and
nonlinear models on input physics. We found that the physical
structure of linear models is marginally affected by the interpolation
methods, based on bicubic splines, we are currently using to estimate
the opacity and its derivatives. Interestingly enough, we also found
that both linear and nonlinear convective models are also marginally
affected by the adopted EOS. We constructed several sequences of
pulsation models at solar chemical composition ($Y=0.28$, and
$Z=0.02$) using the analytical EOS developed by Stellingwerf and the
recent one developed by Irwin. The comparison suggests that the
difference in the pulsation amplitudes as well as in the topology of
the instability strip is negligible.

We selected the Irwin EOS, since it is available in analytical form
and it allows us to change the chemical composition as well as the
abundance of individual elements. To compromise between accuracy and
numerical complexity we computed with the analytical EOS several
tabular EOS by changing the grid resolution in temperature and
density. The comparison between models constructed with analytical and
with the tabular EOS suggests that the difference is marginal once the
step ranges from 0.05 to 0.1 dex in $\log T$ and it is equal to 0.5 dex 
in $\log R$. {\em Note that the use of bicubic spline interpolations both 
for opacity and EOS tables provides the unique opportunity to avoid the
calculation of numerical derivatives. The EOS first and second order
derivatives are estimated by means of the analytical EOS or by means
of analytical derivatives of the interpolating function. The opacity
first order derivatives are evaluated by means of analytical
derivatives of the interpolating function.}

According to recent theoretical investigations we performed several
tests to single out the dependence of pulsation predictions on the ML
relation as well as on the spatial resolution across the H and the He
partial ionization regions. We found that nonlinear models
are marginally affected by ML relations available in the literature.
Note that current ML relations rely on evolutionary prescriptions (CCS; 
ABHA; Bono et al. 2000) that neglect convective core overshooting, 
mass loss, and rotation. 
Both the location and the temperature width of the nonlinear
instability strip present differences on average smaller than 200~K;
while the pulsation amplitudes attain quite similar values.
Otherwise, the increase in the spatial resolution across the partial
ionization regions somehow affects the pulsation properties of
Cepheids. In fact, the instability strip based on finer models moves
by approximately 200-300~K toward cooler effective temperatures when
compared with models based on a coarse zoning. Moreover, fine zoning
models present bolometric amplitudes that are 25\,\% larger than the 
coarse zoning models.

As a whole, the differences between current models with new input
physics and parameters, and predictions by BCM00 marginally affect the
overall trend inside the instability strip, since the slopes of both
blue and red edges predicted by the two sets of models are quite
similar.  This finding is strongly supported by the evidence
that the Period-Radius (PR) relation predicted by current models at
solar chemical composition is the following:

\[
\log R = 1.173 \,(\pm \,0.008) + 0.676 \,(\pm \,0.006) \log P\mathrm{;} 
\] 

\noindent
while the models constructed by BCM00 supply 

\[
\log R = 1.191 \,(\pm \,0.006) + 0.654 \,(\pm \,0.005) \log P\mathrm{,}   
\] 

\noindent
where the radius $R$ is in solar units and the period $P$ in days.  The
difference between these two relations is quite small and ranges
from 0.005 to 0.002 dex when moving from $\log P=0.5$ to $\log P=1.0$.

Moreover and even more importantly, we found that spatial resolution
also affects $P_{HP}$, i.e. the pulsation period at the center of the
Hertzsprung Progression. The new models show that $P_{HP}$ ranges from
9.65 for $M=6.55 \,M_{\sun}$ to 9.84 days for $M=7 \,M_{\sun}$. These
estimates, when compared with models constructed by BCM00, agree
reasonable well with empirical estimates based on light curves
($P_{HP}=10.0\pm 0.5$\,days, \citealt{mbm92}) as well as on radial
velocity curves ($P_{HP}=9.95\pm 0.05$\,days, Moskalik et al. 2000).
Preliminary qualitative results indicate that the new models might
also account for the HP that has been detected in the short-period
range (Kienzle et al. 1999, and references therein).  Current models
do account for the shift of $P_{HP}$ toward shorter periods when
moving toward more metal-rich stellar systems (BCM00, BMS00). However,
more detailed calculations are required to figure out whether current
nonlinear pulsation models do account for the {\em double-peaked}
distribution disclosed by Galactic and Magellanic Bump Cepheids in the
amplitude vs period plane.

To further constrain the zoning effect, we computed several linear and
nonlinear models across the instability strip by increasing the
spatial resolution from 35 to $\approx 50$, i.e. $\Delta T\approx
400$~K. We found that the difference between these two sets of models
is negligible, and indeed the difference among linear periods and
growth rates is smaller than 0.5-1\,\% across the instability strip.  
The relative difference among nonlinear observables is even smaller, 
in particular the shift in temperature of the instability strip is 
smaller than 100~K, while the change in luminosity and velocity 
amplitudes is smaller than $\sim 0.1\,\%$.

The extension of current theoretical framework to Magellanic Cepheids 
seems quite promising not only for the intrinsic properties of 
variable stars but also to figure out whether the input physics,
and/or the spatial resolution of pulsation models do affect the 
topology of the instability strip, and in turn predictions concerning 
the PL and the PLC relations.  
At the same time, we are also interested in performing a more
quantitative comparison between predicted and empirical light curves
using both the Fourier technique (Ngeow et al. 2003) and the principal
component analysis (Kanbur et al. 2002). The main goal of this project
is to constrain the accuracy of current pulsation models, and in turn
to figure out whether the decomposition parameters can be safely
adopted to estimate intrinsic parameters of classical Cepheids.

\acknowledgments 
It is a pleasure to thank A. W. Irwin for detailed information and 
enlightening suggestions concerning the use of the analytical EOS 
he developed. We are also grateful to V. Castellani for several 
enlightening discussions and for a critical reading of an early draft 
of this manuscript. We wish to warmly thank an anonymous referee for 
several comments and suggestions that improved the content and the 
readability of the paper. This work was supported by MIUR/COFIN~2002 
under the project (\#028935): "Stellar Populations in Local Group Galaxies". 

\clearpage

\clearpage 

\clearpage 


\clearpage 

\clearpage 




\clearpage 

\clearpage




\clearpage
\begin{deluxetable}{ccccccccc}
\tablecolumns{9} 
\tablewidth{0pt}
\tablecaption{Relative difference in pulsation periods and growth rates 
for linear models constructed by adopting different input physics.  
\label{t1}}
\tablehead{
\colhead{}&\colhead{}&\colhead{}&\multicolumn{3}{c}{Period}
&\multicolumn{3}{c}{Growth Rate}\\ 
\colhead{$M$\tablenotemark{a}}& 
\colhead{$log L$\tablenotemark{b}}& 
\colhead{$T_e$\tablenotemark{c}}&
\colhead{$\Delta P_{0}/P_{0}$\tablenotemark{d}}&
\colhead{$\Delta P_{1}/P_{1}$}&
\colhead{$\Delta P_{2}/P_{2}$}&
\colhead{$\Delta \eta_{0}/\eta_{0}$\tablenotemark{e}}&
\colhead{$\Delta \eta_{1}/\eta_{1}$}&
\colhead{$\Delta \eta_{2}/\eta_{2}$}}
\startdata
\multicolumn{9}{c}{Test on Opacity\tablenotemark{(1)}}\\
5.0 & 3.07 & 5600 & 0.0003  & 0.0004  &  0.0001 & -0.004 & -0.014 & -0.090\\
9.0 & 3.92 & 4800 & 0.0003  & 0.0004  & -0.0001 &  0.005 & 0.006  & 0.007\\
13.0& 4.46 & 3800 & -0.0001 & -0.0001 & -0.0001 & -0.003 & -0.200 & -0.003\\
\multicolumn{9}{c}{Test on EOS\tablenotemark{(2)}}\\
5.0 & 3.07 & 5600 & -0.0009 & -0.0005 & -0.0003 & 0.003  & 0.003 & 0.0005\\
9.0 & 3.92 & 4800 & 0.003   & 0.002   & 0.001   & -0.006 & 0.003  & -0.010\\
13.0& 4.46 & 3800 &  0.0002 & -0.0005 & -0.0007 & -0.001 & 0.010 & -0.017\\
\multicolumn{9}{c}{Test on EOS\tablenotemark{(3)}}\\
5.0 & 3.07 & 5600 &  0.004 & 0.006  & 0.005  &  0.05 &  0.02 & -0.02\\
9.0 & 3.92 & 4800 & -0.012 & -0.030 & -0.030 & -0.05 & -0.02 & -0.03\\
13.0& 4.46 & 3800 & -0.003 & -0.002 & -0.003 & -0.03 & -0.02 & 0.01\\
\enddata
\tablenotetext{a}{Stellar mass (solar units).\\
\hspace*{2.5mm} $^b$Logarithmic luminosity (solar units).\\
\hspace*{2.5mm} $^c$Static effective temperature (K).\\
\hspace*{2.5mm} $^{d}$Relative difference in periods. $P_{i}$ with
$i=0,1,2$ indicates fundamental, first-overtone, and
second-overtone periods, respectively.\\
\hspace*{2.5mm} $^{e}$Relative difference in growth rates. $\eta_{i}$
with $i=0,1,2$ indicates the fundamental, first-overtone, and
second-overtone growth rates, respectively.\\
\hspace*{2.5mm} $^{(1)}$The linear models were constructed by adopting two 
different methods to interpolate the opacity tables (see \S~\ref{opacity}).\\
\hspace*{2.5mm} $^{(2)}$The linear models were constructed by adopting the 
analytical Irwin EOS and the tabular form of the same EOS (see \S~\ref{eos}).\\
\hspace*{2.5mm} $^{(3)}$The linear models were constructed by adopting the
analytical Stellingwerf EOS and the tabular Irwin EOS (see \S~\ref{lin1}).}
\end{deluxetable}
\clearpage

\begin{deluxetable}{cccccccccc}
\tablecolumns{10} 
\tablewidth{0pt} 
\tablecaption{Input parameters and nonlinear observables for first-overtone 
and fundamental Cepheid models at solar chemical composition.\label{t2}}
\tablehead{
\colhead{$M$\tablenotemark{a}}&
\colhead{$log L$\tablenotemark{b}}&
\colhead{$T_e$\tablenotemark{c}}&
\colhead{$P$\tablenotemark{d}}& 
\colhead{$\log{\overline{R}}$\tablenotemark{e}}& 
\colhead{$\Delta R/R_{ph}$\tablenotemark{f}}& 
\colhead{$\Delta u$\tablenotemark{g}}&
\colhead{$\Delta M_{bol}$\tablenotemark{h}}& 
\colhead{$\Delta {\log{g_s}}$\tablenotemark{i}}&  
\colhead{$\Delta T_{e}$\tablenotemark{j}}\\
\colhead{(1)}&
\colhead{(2)}&
\colhead{(3)}&
\colhead{(4)}&
\colhead{(5)}&
\colhead{(6)}&
\colhead{(7)}&
\colhead{(8)}& 
\colhead{(9)}& 
\colhead{(10)}}
\startdata
\multicolumn{10}{c}{First-overtone}\\
5.0 & 3.07  &   6100  &  2.0565   &  1.488  &  0.045 &   34.87  &  0.497  &  0.04  &    740 \\
5.0 & 3.07  &   6000  &  2.1733   &  1.502  &  0.051 &   39.27  &  0.505  &  0.04  &    728 \\
5.0 & 3.07  &   5900  &  2.2958   &  1.516  &  0.050 &   37.94  &  0.438  &  0.04  &    620 \\
5.0 & 3.07  &   5800  &  2.4253   &  1.531  &  0.044 &   32.38  &  0.341  &  0.04  &    479 \\
\multicolumn{10}{c}{Fundamental}\\			
5.0 & 3.07  &   5700  &  3.6840   &  1.548  &  0.115 &   57.88  &  0.742  &  0.10  &   1037 \\
5.0 & 3.07  &   5600  &  3.8991   &  1.564  &  0.120 &   57.55  &  0.656  &  0.11  &    906 \\
5.0 & 3.07  &   5500  &  4.1422   &  1.578  &  0.105 &   50.63  &  0.511  &  0.09  &    712 \\
5.0 & 3.07  &   5400  &  4.3949   &  1.593  &  0.094 &   43.80  &  0.396  &  0.08  &    610 \\
5.0 & 3.07  &   5300  &  4.6688   &  1.606  &  0.074 &   33.22  &  0.288  &  0.07  &    445 \\
5.0 & 3.07  &   5200  &  4.9913   &  1.622  &  0.038 &   16.42  &  0.127  &  0.03  &    212 \\
6.55& 3.46  &   5600  &   7.0730  &  1.758  &  0.072 &   31.13  &  0.509  &  0.06  &    703 \\
6.55& 3.46  &   5500  &   7.4988  &  1.773  &  0.079 &   34.83  &  0.524  &  0.07  &    712 \\
6.55& 3.46  &   5400  &   7.9803  &  1.790  &  0.077 &   34.24  &  0.483  &  0.07  &    643 \\
6.55& 3.46  &   5300  &   8.4829  &  1.804  &  0.074 &   32.45  &  0.412  &  0.06  &    554 \\
6.55& 3.46  &   5200  &   9.0525  &  1.820  &  0.059 &   23.98  &  0.258  &  0.05  &    375 \\
6.55& 3.46  &   5150  &   9.3328  &  1.828  &  0.048 &   18.40  &  0.211  &  0.04  &    287 \\
6.55& 3.46  &   5100  &   9.6496  &  1.836  &  0.042 &   15.39  &  0.188  &  0.04  &    223 \\
6.55& 3.46  &   5050  &   9.9560  &  1.845  &  0.047 &   18.58  &  0.212  &  0.04  &    269 \\
6.55& 3.46  &   5000  &  10.3025  &  1.853  &  0.060 &   24.68  &  0.266  &  0.05  &    354 \\
6.55& 3.46  &   4900  &  11.0464  &  1.869  &  0.076 &   29.66  &  0.298  &  0.07  &    418 \\
7.0 & 3.56  &   5500  &   8.6976  &  1.821  &  0.057 &   23.71  &  0.372  &  0.05  &    521 \\
7.0 & 3.56  &   5400  &   9.2348  &  1.838  &  0.064 &   26.40  &  0.377  &  0.06  &    532 \\
7.0 & 3.56  &   5300  &   9.8443  &  1.853  &  0.057 &   21.74  &  0.282  &  0.05  &    394 \\
7.0 & 3.56  &   5200  &  10.4821  &  1.869  &  0.064 &   26.03  &  0.341  &  0.06  &    434 \\
7.0 & 3.56  &   5100  &  11.2346  &  1.888  &  0.110 &   48.21  &  0.600  &  0.10  &    798 \\ 
7.0 & 3.56  &   5000  &  12.0295  &  1.908  &  0.149 &   63.24  &  0.762  &  0.13  &   1029 \\ 
7.0 & 3.56  &   4900  &  12.8869  &  1.926  &  0.171 &   68.79  &  0.828  &  0.15  &   1128 \\
7.0 & 3.56  &   4800  &  13.7777  &  1.941  &  0.170 &   62.63  &  0.723  &  0.15  &   1038 \\ 
9.0 & 3.92  &   5200  &   18.7620  &  1.951  &  0.198 &   64.96  &  1.180  &  0.17  &   1721 \\
9.0 & 3.92  &   5100  &   20.0704  &  2.079  &  0.220 &   71.09  &  1.227  &  0.19  &   1751 \\
9.0 & 3.92  &   4900  &   22.9589  &  2.117  &  0.218 &   67.38  &  0.922  &  0.19  &   1283 \\
9.0 & 3.92  &   4800  &   24.5763  &  2.129  &  0.204 &   59.74  &  0.750  &  0.18  &   1116 \\
9.0 & 3.92  &   4600  &   28.2280  &  2.158  &  0.174 &   43.20  &  0.490  &  0.15  &    800 \\
9.0 & 3.92  &   4500  &   30.2840  &  2.175  &  0.165 &   39.88  &  0.415  &  0.15  &    667 \\
9.0 & 3.92  &   4400  &   32.4571  &  2.188  &  0.110 &   24.33  &  0.236  &  0.10  &    363 \\
11.0& 4.21  &   5000  &   34.2116  &  2.236  &  0.232 &   36.29  &  0.641  &  0.21  &    901 \\
11.0& 4.21  &   4900  &   36.7837  &  2.259  &  0.288 &   45.70  &  0.832  &  0.27  &   1128 \\
11.0& 4.21  &   4800  &   39.4572  &  2.276  &  0.329 &   50.28  &  0.879  &  0.31  &   1128 \\
11.0& 4.21  &   4600  &   45.5276  &  2.311  &  0.387 &   52.14  &  0.804  &  0.38  &    918 \\
11.0& 4.21  &   4500  &   48.9255  &  2.326  &  0.408 &   49.24  &  0.725  &  0.41  &    845 \\
11.0& 4.21  &   4300  &   56.4072  &  2.357  &  0.436 &   40.54  &  0.545  &  0.46  &    682 \\
11.0& 4.21  &   4200  &   60.4930  &  2.371  &  0.441 &   34.06  &  0.444  &  0.49  &    564 \\
11.0& 4.21  &   4100  &   64.9105  &  2.385  &  0.438 &   25.57  &  0.310  &  0.50  &    416 \\
11.0& 4.21  &   4000  &   69.7311  &  2.395  &  0.405 &    7.14  &  0.084  &  0.49  &    106 \\
13.0& 4.46  &   4800  &   58.9766  &  2.393  &  0.480 &   31.59  &  0.581  &  0.52  &    775 \\
13.0& 4.46  &   4700  &   63.2735  &  2.412  &  0.521 &   39.56  &  0.767  &  0.57  &    942 \\
13.0& 4.46  &   4500  &   73.6824  &  2.447  &  0.565 &   43.08  &  0.742  &  0.65  &    806 \\
13.0& 4.46  &   4400  &   78.8145  &  2.464  &  0.577 &   43.43  &  0.729  &  0.68  &    746 \\
13.0& 4.46  &   4200  &   91.4053  &  2.496  &  0.588 &   35.29  &  0.591  &  0.73  &    597 \\
13.0& 4.46  &   4100  &   97.9617  &  2.509  &  0.587 &   30.92  &  0.524  &  0.76  &    539 \\
13.0& 4.46  &   3900  &  113.0619  &  2.537  &  0.572 &   18.45  &  0.293  &  0.80  &    326 \\
13.0& 4.46  &   3800  &  120.3557  &  2.549  &  0.556 &   12.03  &  0.197  &  0.81  &    220 \\
13.0& 4.46  &   3700  &  127.1189  &  2.559  &  0.535 &    6.66  &  0.111  &  0.82  &    109 \\
\enddata
\tablenotetext{a}{Stellar mass (solar units).\\  
\hspace*{2.5mm} $^b$Logarithmic luminosity (solar units).\\  
\hspace*{2.5mm} $^c$Effective temperature (K).\\
\hspace*{2.5mm} $^d$Period (day).\\
\hspace*{2.5mm} $^e$Logarithmic mean radius (solar units).\\ 
\hspace*{2.5mm} $^f$Fractional radius variation.\\  
\hspace*{2.5mm} $^g$Radial velocity amplitude (Km~s$^{-1}$).\\
\hspace*{2.5mm} $^h$Bolometric amplitude (mag).\\
\hspace*{2.5mm} $^i$Amplitude of logarithmic static gravity.\\  
\hspace*{2.5mm} $^j$Effective temperature variation (K).}
\end{deluxetable}
\clearpage

\begin{deluxetable}{cccccccccc}
\tablecolumns{10} 
\tablewidth{0pt} 
\tablecaption{Same as Table~\ref{t2}, but the models were constructed 
by adopting the Stellingwerf EOS, the ML relation by Bono et al. (2000), 
and a fine zoning across Hydrogen and Helium ionization regions.\label{t3}}
\tablehead{
\colhead{$M$\tablenotemark{a}}&
\colhead{$log L$\tablenotemark{b}}&
\colhead{$T_e$\tablenotemark{c}}&
\colhead{$P$\tablenotemark{d}}& 
\colhead{$\log{\overline{R}}$\tablenotemark{e}}& 
\colhead{$\Delta R/R_{ph}$\tablenotemark{f}}& 
\colhead{$\Delta u$\tablenotemark{g}}&
\colhead{$\Delta M_{bol}$\tablenotemark{h}}& 
\colhead{$\Delta {\log{g_s}}$\tablenotemark{i}}&  
\colhead{$\Delta T_{e}$\tablenotemark{j}}\\
\colhead{(1)}&
\colhead{(2)}&
\colhead{(3)}&
\colhead{(4)}&
\colhead{(5)}&
\colhead{(6)}&
\colhead{(7)}&
\colhead{(8)}& 
\colhead{(9)}& 
\colhead{(10)}}
\startdata
\multicolumn{10}{c}{First-overtone}\\
5.0 & 3.07  &   6200  &    1.9373  &  1.472  &  0.038 &   30.12  &  0.462  &  0.03  &    706 \\
5.0 & 3.07  &   6100  &    2.0451  &  1.488  &  0.056 &   44.00  &  0.626  &  0.05  &    932 \\
5.0 & 3.07  &   6000  &    2.1559  &  1.502  &  0.057 &   45.38  &  0.583  &  0.05  &    843 \\
5.0 & 3.07  &   5900  &    2.2745  &  1.517  &  0.054 &   42.07  &  0.489  &  0.05  &    682 \\
5.0 & 3.07  &   5800  &    2.4248  &  1.531  &  0.048 &   36.05  &  0.381  &  0.04  &    526 \\
\multicolumn{10}{c}{Fundamental}\\

5.0 & 3.07  &   5700  &    3.6675  &  1.548  &  0.116 &   59.33  &  0.762  &  0.10  &   1068 \\
5.0 & 3.07  &   5600  &    3.8855  &  1.563  &  0.113 &   55.74  &  0.634  &  0.10  &    862 \\
5.0 & 3.07  &   5500  &    4.1284  &  1.578  &  0.105 &   50.78  &  0.522  &  0.09  &    724 \\
5.0 & 3.07  &   5400  &    4.3816  &  1.592  &  0.094 &   43.88  &  0.406  &  0.08  &    617 \\
5.0 & 3.07  &   5300  &    4.6624  &  1.606  &  0.074 &   33.33  &  0.297  &  0.07  &    459 \\
5.0 & 3.07  &   5200  &    4.9612  &  1.622  &  0.042 &   17.99  &  0.143  &  0.04  &    238 \\
6.55& 3.46  &   5700  &    6.6665  &  1.742  &  0.026 &   10.98  &  0.186  &  0.02  &    269\\
6.55& 3.46  &   5600  &    7.0588  &  1.757  &  0.073 &   32.08  &  0.521  &  0.06  &    720\\
6.55& 3.46  &   5500  &    7.5039  &  1.774  &  0.079 &   35.07  &  0.533  &  0.07  &    725\\
6.55& 3.46  &   5400  &    7.9632  &  1.790  &  0.077 &   34.43  &  0.490  &  0.07  &    657\\
6.55& 3.46  &   5300  &    8.5037  &  1.804  &  0.073 &   31.50  &  0.399  &  0.06  &    548\\ 
6.55& 3.46  &   5200  &    9.0372  &  1.819  &  0.056 &   22.64  &  0.255  &  0.05  &    365\\
6.55& 3.46  &   5150  &    9.3354  &  1.828  &  0.046 &   17.18  &  0.213  &  0.04  &    275\\
6.55& 3.46  &   5100  &    9.6520  &  1.836  &  0.045 &   17.22  &  0.210  &  0.04  &    259\\
6.55& 3.46  &   5050  &    9.9832  &  1.845  &  0.054 &   22.29  &  0.257  &  0.05  &    334\\
6.55& 3.46  &   5000  &   10.3117  &  1.854  &  0.071 &   29.84  &  0.333  &  0.06  &    446\\
6.55& 3.46  &   4900  &   11.0254  &  1.871  &  0.106 &   43.62  &  0.474  &  0.09  &    657\\
6.55& 3.46  &   4800  &   11.7725  &  1.886  &  0.056 &   19.61  &  0.179  &  0.05  &    271\\
7.0 & 3.56  &   5600  &    8.1908  &  1.804  &  0.022 &    9.01  &  0.150  &  0.02  &    217 \\
7.0 & 3.56  &   5500  &    8.7064  &  1.822  &  0.057 &   23.74  &  0.371  &  0.05  &    528 \\
7.0 & 3.56  &   5400  &    9.2405  &  1.837  &  0.058 &   23.42  &  0.330  &  0.05  &    475 \\
7.0 & 3.56  &   5300  &    9.8539  &  1.853  &  0.054 &   19.83  &  0.284  &  0.05  &    349 \\
7.0 & 3.56  &   5200  &   10.5189  &  1.869  &  0.079 &   33.87  &  0.449  &  0.07  &    586 \\
7.0 & 3.56  &   5100  &   11.2733  &  1.888  &  0.125 &   55.06  &  0.711  &  0.11  &    942 \\ 
7.0 & 3.56  &   5000  &   12.0393  &  1.908  &  0.157 &   66.96  &  0.837  &  0.14  &   1118 \\
7.0 & 3.56  &   4900  &   12.8744  &  1.927  &  0.174 &   69.36  &  0.865  &  0.15  &   1181 \\
7.0 & 3.56  &   4800  &   13.8014  &  1.943  &  0.180 &   65.97  &  0.799  &  0.16  &   1135 \\ 
7.0 & 3.56  &   4700  &   14.6685  &  1.951  &  0.023 &    7.17  &  0.064  &  0.02  &    100 \\
9.0 & 3.92  &   5200  &   18.7266  &  2.062  &  0.200 &   66.06  &  1.217  &  0.17  &   1778 \\
9.0 & 3.92  &   5100  &   20.0248  &  2.080  &  0.222 &   72.40  &  1.292  &  0.19  &   1854 \\
9.0 & 3.92  &   4900  &   22.9230  &  2.113  &  0.222 &   69.25  &  1.004  &  0.20  &   1382 \\
9.0 & 3.92  &   4800  &   24.4934  &  2.130  &  0.212 &   62.75  &  0.822  &  0.19  &   1150 \\
9.0 & 3.92  &   4600  &   28.1603  &  2.159  &  0.183 &   46.07  &  0.534  &  0.16  &    846 \\
9.0 & 3.92  &   4500  &   30.2792  &  2.177  &  0.180 &   43.81  &  0.476  &  0.16  &    740 \\
9.0 & 3.92  &   4400  &   32.4045  &  2.190  &  0.154 &   36.54  &  0.378  &  0.14  &    573 \\
9.0 & 3.92  &   4300  &   34.7587  &  2.202  &  0.115 &   10.86  &  0.104  &  0.11  &    153 \\
11.0& 4.21  &   4900  &   36.5833  &  2.257  &  0.282 &   43.12  &  0.787  &  0.26  &   1072 \\ 
11.0& 4.21  &   4800  &   39.2919  &  2.274  &  0.325 &   48.35  &  0.876  &  0.31  &   1125  \\
11.0& 4.21  &   4600  &   45.3609  &  2.311  &  0.385 &   50.60  &  0.791  &  0.38  &    909  \\
11.0& 4.21  &   4500  &   48.7902  &  2.326  &  0.407 &   48.44  &  0.721  &  0.41  &    826  \\
11.0& 4.21  &   4300  &   56.2860  &  2.357  &  0.438 &   40.30  &  0.540  &  0.47  &    677  \\
11.0& 4.21  &   4200  &   60.5200  &  2.372  &  0.444 &   34.31  &  0.443  &  0.49  &    565  \\
11.0& 4.21  &   4100  &   64.9298  &  2.386  &  0.441 &   25.92  &  0.309  &  0.51  &    415  \\
11.0& 4.21  &   4000  &   69.7022  &  2.397  &  0.415 &   10.08  &  0.108  &  0.50  &    150  \\
13.0& 4.46  &   4800  &   58.3580  &  2.389  &  0.437 &   13.56  &  0.224  &  0.49  &    314 \\
13.0& 4.46  &   4700  &   63.0748  &  2.411  &  0.516 &   37.00  &  0.703  &  0.57  &    881 \\
13.0& 4.46  &   4500  &   73.1909  &  2.449  &  0.564 &   42.79  &  0.752  &  0.65  &    817  \\
13.0& 4.46  &   4400  &   78.6635  &  2.465  &  0.578 &   42.26  &  0.714  &  0.68  &    734 \\
13.0& 4.46  &   4200  &   90.9012  &  2.496  &  0.591 &   36.73  &  0.604  &  0.74  &    586  \\
13.0& 4.46  &   4100  &   98.2004  &  2.512  &  0.590 &   30.99  &  0.512  &  0.76  &    518  \\
13.0& 4.46  &   4000  &  104.9605  &  2.525  &  0.584 &   24.93  &  0.392  &  0.78  &    420  \\
13.0& 4.46  &   3900  &  112.4829  &  2.537  &  0.574 &   18.75  &  0.285  &  0.80  &    316  \\
13.0& 4.46  &   3800  &  119.3546  &  2.548  &  0.557 &   13.01  &  0.198  &  0.81  &    223  \\
13.0& 4.46  &   3700  &  126.5626  &  2.558  &  0.533 &    5.66  &  0.081  &  0.82  &     92  \\
\enddata
\tablenotetext{a}{Stellar mass (solar units).\\  
\hspace*{2.5mm} $^b$Logarithmic luminosity (solar units).\\  
\hspace*{2.5mm} $^c$Effective temperature (K).\\
\hspace*{2.5mm} $^d$Period (day).\\
\hspace*{2.5mm} $^e$Logarithmic mean radius (solar units).\\ 
\hspace*{2.5mm} $^f$Fractional radius variation.\\  
\hspace*{2.5mm} $^g$Radial velocity amplitude (Km~s$^{-1}$).\\
\hspace*{2.5mm} $^h$Bolometric amplitude (mag).\\
\hspace*{2.5mm} $^i$Amplitude of logarithmic static gravity.\\  
\hspace*{2.5mm} $^j$Effective temperature variation (K).}
\end{deluxetable}
\clearpage

\begin{deluxetable}{cccccccccc}
\tablecolumns{10} 
\tablewidth{0pt} 
\tablecaption{Input parameters and nonlinear observables for Cepheid
models with $M=11\,M_{\sun}$ and solar chemical composition. These 
models were constructed by adopting the Irwin EOS, the ML relation used 
by BCM00, and a fine zoning across Hydrogen and Helium ionization 
regions.\label{t4}}
\tablehead{
\colhead{$M$\tablenotemark{a}}&
\colhead{$log L$\tablenotemark{b}}&
\colhead{$T_e$\tablenotemark{c}}&
\colhead{$P$\tablenotemark{d}}& 
\colhead{$\log{\overline{R}}$\tablenotemark{e}}& 
\colhead{$\Delta R/R_{ph}$\tablenotemark{f}}& 
\colhead{$\Delta u$\tablenotemark{g}}&
\colhead{$\Delta M_{bol}$\tablenotemark{h}}& 
\colhead{$\Delta {\log{g_s}}$\tablenotemark{i}}&  
\colhead{$\Delta T_{e}$\tablenotemark{j}}\\
\colhead{(1)}&
\colhead{(2)}&
\colhead{(3)}&
\colhead{(4)}&
\colhead{(5)}&
\colhead{(6)}&
\colhead{(7)}&
\colhead{(8)}& 
\colhead{(9)}& 
\colhead{(10)}}
\startdata
11.0& 4.40  &   4800  &   59.3229  &  2.364  &  0.446 &   32.48  &  0.632  &  0.47  &    827\\ 
11.0& 4.40  &   4600  &   68.5505  &  2.401  &  0.507 &   41.51  &  0.806  &  0.55  &    916\\
11.0& 4.40  &   4500  &   73.7417  &  2.418  &  0.525 &   42.27  &  0.779  &  0.59  &    828\\
11.0& 4.40  &   4300  &   85.2422  &  2.449  &  0.546 &   37.21  &  0.663  &  0.64  &    649\\
11.0& 4.40  &   4200  &   91.7339  &  2.465  &  0.550 &   32.57  &  0.578  &  0.67  &    579\\
11.0& 4.40  &   4100  &   98.6821  &  2.479  &  0.549 &   26.80  &  0.462  &  0.69  &    477\\
11.0& 4.40  &   4000  &  105.8632  &  2.494  &  0.546 &   22.17  &  0.392  &  0.71  &    397\\
11.0& 4.40  &   3900  &  113.1294  &  2.506  &  0.532 &   15.18  &  0.283  &  0.73  &    295\\
\enddata
\tablenotetext{a}{Stellar mass (solar units).\\  
\hspace*{2.5mm} $^b$Logarithmic luminosity (solar units).\\  
\hspace*{2.5mm} $^c$Effective temperature (K).\\
\hspace*{2.5mm} $^d$Period (day).\\
\hspace*{2.5mm} $^e$Logarithmic mean radius (solar units).\\ 
\hspace*{2.5mm} $^f$Fractional radius variation.\\  
\hspace*{2.5mm} $^g$Radial velocity amplitude (Km~s$^{-1}$).\\
\hspace*{2.5mm} $^h$Bolometric amplitude (mag).\\
\hspace*{2.5mm} $^i$Amplitude of logarithmic static gravity.\\  
\hspace*{2.5mm} $^j$Effective temperature variation (K).}
\end{deluxetable}
\clearpage

\begin{deluxetable}{cccccccccc}
\tablecolumns{10} 
\tablewidth{0pt} 
\tablecaption{Input parameters and nonlinear observables for Cepheid
models of $M=11\,M_{\sun}$ and solar chemical composition. These models 
were constructed by adopting the Irwin EOS, the ML relation by 
Bono et al. (2000), and a coarse zoning across Hydrogen and Helium 
ionization regions. \label{t5}}
\tablehead{
\colhead{$M$\tablenotemark{a}}&
\colhead{$log L$\tablenotemark{b}}&
\colhead{$T_e$\tablenotemark{c}}&
\colhead{$P$\tablenotemark{d}}& 
\colhead{$\log{\overline{R}}$\tablenotemark{e}}& 
\colhead{$\Delta R/R_{ph}$\tablenotemark{f}}& 
\colhead{$\Delta u$\tablenotemark{g}}&
\colhead{$\Delta M_{bol}$\tablenotemark{h}}& 
\colhead{$\Delta {\log{g_s}}$\tablenotemark{i}}&  
\colhead{$\Delta T_{e}$\tablenotemark{j}}\\
\colhead{(1)}&
\colhead{(2)}&
\colhead{(3)}&
\colhead{(4)}&
\colhead{(5)}&
\colhead{(6)}&
\colhead{(7)}&
\colhead{(8)}& 
\colhead{(9)}& 
\colhead{(10)}}
\startdata
11.0& 4.21  & 4900  &   36.9506  &  2.257  &  0.282 &   42.13  &  0.725  &  0.26  &    984\\ 
11.0& 4.21  & 4800  &   39.6741  &  2.275  &  0.321 &   46.38  &  0.760  &  0.30  &    987\\
11.0& 4.21  & 4600  &   45.7762  &  2.309  &  0.376 &   44.98  &  0.648  &  0.37  &    835\\
11.0& 4.21  & 4500  &   49.0666  &  2.325  &  0.396 &   42.01  &  0.579  &  0.40  &    763\\
11.0& 4.21  & 4300  &   56.7848  &  2.354  &  0.417 &   30.17  &  0.382  &  0.45  &    532\\
11.0& 4.21  & 4200  &   60.9255  &  2.368  &  0.408 &   18.66  &  0.216  &  0.46  &    309\\
\enddata
\tablenotetext{a}{Stellar mass (solar units).\\  
\hspace*{2.5mm} $^b$Logarithmic luminosity (solar units).\\  
\hspace*{2.5mm} $^c$Effective temperature (K).\\
\hspace*{2.5mm} $^d$Period (day).\\
\hspace*{2.5mm} $^e$Logarithmic mean radius (solar units).\\ 
\hspace*{2.5mm} $^f$Fractional radius variation.\\  
\hspace*{2.5mm} $^g$Radial velocity amplitude (Km~s$^{-1}$).\\
\hspace*{2.5mm} $^h$Bolometric amplitude (mag).\\
\hspace*{2.5mm} $^i$Amplitude of logarithmic static gravity.\\  
\hspace*{2.5mm} $^j$Effective temperature variation (K).}
\end{deluxetable}
\clearpage 

\begin{deluxetable}{cccccccccc}
\tablecolumns{10} 
\tablewidth{0pt} 

\tablecaption{Input parameters and nonlinear observables for fundamental 
Cepheid models with $M=6.55\,M_{\sun}$ and solar chemical composition. 
These models were constructed by adopting the Stellingwerf EOS, 
as well as the ML relation and the zoning used by BCM00.\label{t6}}

\tablehead{
\colhead{$M$\tablenotemark{a}}&
\colhead{$log L$\tablenotemark{b}}&
\colhead{$T_e$\tablenotemark{c}}&
\colhead{$P$\tablenotemark{d}}& 
\colhead{$\log{\overline{R}}$\tablenotemark{e}}& 
\colhead{$\Delta R/R_{ph}$\tablenotemark{f}}& 
\colhead{$\Delta u$\tablenotemark{g}}&
\colhead{$\Delta M_{bol}$\tablenotemark{h}}& 
\colhead{$\Delta {\log{g_s}}$\tablenotemark{i}}&  
\colhead{$\Delta T_{e}$\tablenotemark{j}}\\
\colhead{(1)}&
\colhead{(2)}&
\colhead{(3)}&
\colhead{(4)}&
\colhead{(5)}&
\colhead{(6)}&
\colhead{(7)}&
\colhead{(8)}& 
\colhead{(9)}& 
\colhead{(10)}}
\startdata
6.55& 3.48  &   5600  &  7.4046   & 1.767 & 0.041 &  16.87 & 0.267 & 0.168 & 300\\
6.55& 3.48  &   5500  &  7.8451   & 1.783 & 0.057 &  24.06 & 0.353 & 0.255 & 400\\
6.55& 3.48  &   5400  &  8.3426   & 1.798 & 0.057 &  23.61 & 0.312 & 0.267 & 350 \\
6.55& 3.48  &   5350  &  8.6278   & 1.806 & 0.053 &  21.14 & 0.262 & 0.251& 300 \\
6.55& 3.48  &   5300  &  8.8797   & 1.814 & 0.047 &  17.98 & 0.218 & 0.228& 250 \\
6.55& 3.48  &   5250  &  9.1559   & 1.822 & 0.043 &  15.55 & 0.204 & 0.210& 200 \\
6.55& 3.48  &   5200  &  9.4667   & 1.830 & 0.045 &  17.03 & 0.217 & 0.216& 250  \\
6.55& 3.48  &   5100  &  10.139   & 1.847 & 0.067 &  26.58 & 0.310 & 0.296& 350 \\
6.55& 3.48  &   5000  &  10.789   & 1.863 & 0.086 &  33.01 & 0.363 & 0.348& 400 \\
\enddata
\tablenotetext{a}{Stellar mass (solar units).\\  
\hspace*{2.5mm} $^b$Logarithmic luminosity (solar units).\\  
\hspace*{2.5mm} $^c$Effective temperature (K).\\
\hspace*{2.5mm} $^d$Period (day).\\
\hspace*{2.5mm} $^e$Logarithmic mean radius (solar units).\\ 
\hspace*{2.5mm} $^f$Fractional radius variation.\\  
\hspace*{2.5mm} $^g$Radial velocity amplitude (Km~s$^{-1}$).\\
\hspace*{2.5mm} $^h$Bolometric amplitude (mag).\\
\hspace*{2.5mm} $^i$Amplitude of logarithmic static gravity.\\  
\hspace*{2.5mm} $^j$Effective temperature variation (K).}
\end{deluxetable}

\end{document}